\documentclass[12pt]{article}
\usepackage{latexsym,amsfonts}
\usepackage{color}
\usepackage{amsmath}
\usepackage{amssymb}
\usepackage{epsfig}
\usepackage{bm}
\usepackage{subfigure}
\begin{document}

\title{\bf \LARGE Dynamics of multi-kinks in the presence of wells and
barriers}
\author{
Stephen W. Goatham{\thanks{E-mail: {\tt swg3@kent.ac.uk}}}~,
Lucy E. Mannering{\thanks{E-mail: {\tt lm231@kent.ac.uk}}}~,\\
Rebecca Hann{\thanks{E-mail: {\tt rebeccahann@tiscali.co.uk}}}
~and 
Steffen Krusch\thanks{E-mail: {\tt S.Krusch@kent.ac.uk}}\\ \\[5pt]
{\normalsize {\sl School of Mathematics, Statistics and Actuarial
Science}}\\
{\normalsize {\sl University of Kent,
Canterbury CT2 7NF, United Kingdom}}
}

\date{July 15, 2010}
\maketitle

\begin{abstract}
Sine-Gordon kinks are a much studied integrable system that possesses
multi-soliton solutions. Recent studies on sine-Gordon kinks with
space-dependent square-well-type potentials have revealed
interesting dynamics of a single kink interacting with wells and barriers. 
In this paper, we study a class of smooth space-dependent potentials and
discuss the dynamics of one kink in the presence of different
wells. We also present values for the critical velocity for different 
types of barriers. Furthermore, we study two kinks interacting with 
various wells and describe interesting trajectories such as 
double-trapping, kink knock-out and double-escape.
\end{abstract}

\newpage
\section{Introduction}
\label{intro}
Topological solitons arise in many classical field theories
\cite{Manton:2004tk}.  One of the simplest systems to admit solitons is 
the sine-Gordon model.  These solitons are known as kinks and occur in 
(1+1) dimensions. The model is described by the Lagrangian density
\begin{equation}
\label{lagrange}
{\cal L} =\frac{1}{2}\left(\frac{\partial \phi}{\partial t}\right)^{2}
-\frac{1}{2}\left(\frac{\partial \phi}{\partial
x}\right)^{2}-\lambda\left(1-\cos{\phi}\right),
\end{equation}
where $\phi(x,t)$ is a scalar field and $\lambda$ is a coupling constant.
Applying the standard variational
principle leads to the field equation
\begin{equation}
\label{SGeq}
\frac{\partial^{2} \phi}{\partial t^{2}}-\frac{\partial^{2} \phi}{\partial
x^{2}}+\lambda \sin{\phi}=0.
\end{equation}
This so-called sine-Gordon equation has applications in a large number of
areas of physical and bio-physical interest, including the Josephson
effect \cite{Josephson}, nuclear physics \cite{Perring:1962vs}, non-linear
optics \cite{McCall}, ferromagnetic spin chains \cite{Wysin} and wave
propagation in brain microtubules \cite{Abdalla}, amongst others. Taking a
spin chain in the easy plane model of a ferromagnet as an example,
$\lambda$ is determined by the strength of the magnetic field, which
occurs in a direction in the ($x$-$y$) easy plane (see \cite{Wysin} for
details).  When this system becomes inhomogeneous, that is the magnetic 
field has a $z$-dependence, then the corresponding parameter $\lambda$ is 
also space-dependent. 
The Bloch equation describing the system still leads to a
sine-Gordon equation, and solitons can still occur.  We will not discuss
physical interpretations further. In this paper, we shall be interested in
the sine-Gordon model with parameter $\lambda$, which, as a function of
space, is of the form of a smooth well or barrier.  For our system,
$\lambda$ will be positive and, far away from the origin, very close to
unity.

The interaction of a sine-Gordon soliton with a potential obstacle was
first investigated by Fei et al \cite{Fei}.  The authors investigated a 
kink incident on a point defect and found a novel behaviour in which, for
certain incoming velocities, the kink is reflected backwards, due to its
interaction with the defect.  Many of the results in \cite{Fei} were
explained by Goodman and Haberman \cite{Goodman} in terms of a two-bounce
resonance model.  In \cite{Piette:2006gw} Piette and Zakrzewski
investigated a sine-Gordon kink interacting with a potential with
a space-dependent term in the form of a rectangular well.  They found that
back-reflection for some incoming velocities also arises for this system
and gave two effective models to account for this phenomenon. 
Soliton interactions with rectangular wells have also been discussed for 
various other soliton systems \cite{AlAlawi:2007th, AlAlawi:2008aa, 
AlAlawi:2009nr, AlAlawi:2009rt, Ferreira:2007ue, 
Ferreira:2007gu, Piette:2005wz}.
A kink interacting with a smooth well is discussed in 
\cite{Javidan:2006js} in terms of the sine-Gordon model and in \cite 
{Kalbermann:1998pn} for the $\lambda \phi^4$ model.  While these papers 
address the dynamics of a kink interacting with a well, multi-kinks 
interacting with wells have so far not been investigated.

In this paper, section \ref{SGintro} gives a review of sine-Gordon kinks.  
Then section \ref{potentials} introduces a two-parameter family of
space-dependent potentials which include a variety of wells and barriers.
In section \ref{onekink} the results of simulations of the dynamics of a
single kink interacting with a particular well are given. Following this, 
in section \ref{twokinks}, the dynamics of a moving kink
incident on a well with another initially at rest in the well are
presented.  The paper also gives plots of the scattering data for
$1$-kink and $2$-kink systems with two different types of wells, these are
discussed at the end of section \ref{twokinks}.
The paper ends with a conclusion.

\section{Sine-Gordon kinks}
\label{SGintro}
In this section, we recall some basic facts about sine-Gordon kinks and
set up our notation.

The energy of static sine-Gordon kink is
\begin{equation}
\label{E}
E=\int\limits_{-\infty}^{\infty}\left(\frac{1}{2}\left(\frac{d \phi}{d
x}\right)^{2}+\lambda\left(1-\cos{\phi}\right)\right)\, {\rm d}x.
\end{equation}
One can also define a super potential, $W,$ given by
\begin{equation}
\label{super}
\frac{1}{2}\left(\frac{d W}{d \phi}\right)^{2}=\lambda
\left(1-\cos{\phi}\right).
\end{equation}
The energy (\ref{E}) can then be rewritten in terms of two integrals
\begin{equation}
E=\frac{1}{2}\int\limits_{-\infty}^{\infty}\left(\frac{d \phi}{dx} \mp
\frac{dW}{d \phi}\right)^{2}\, {\rm d}x\pm
\int\limits_{\phi(-\infty)}^{\phi(\infty)}\, {\rm d}W.
\end{equation}
For a minimum energy soliton, we obtain the Bogomolny equations
\begin{equation}
\label{Bog}
\frac{d \phi}{dx}= \pm \frac{dW}{d \phi},
\end{equation}
where the $\pm$ sign gives a kink or anti-kink.  Substituting the solution
of (\ref {super}) into (\ref{Bog}) gives the kink field
\begin{equation}
\label{sgkink}
\phi(x) = 4 \arctan\left(\exp\left(\sqrt{\lambda}(x-x_{0})\right)\right),
\end{equation}
where $x_{0}$ is a constant of integration and corresponds to the 
position of the kink. The energy can then be evaluated to be 
$8\sqrt{\lambda},$ so one has 
the Bogomolny bound
\begin{equation}
E\ge 8\sqrt{\lambda}.
\end{equation}
Since the Lagrangian density (\ref{lagrange}) is Lorentz invariant, a
boost can be applied to (\ref{sgkink}) to give a moving kink with field
configuration
\begin{equation}
\label{initialkink}
\phi=4\arctan\left(\exp\left(\gamma\left(x-vt-x_{0}\right)\right)\right),
\end{equation}
where $v$ is the velocity of the kink, with $-1<v<1$, and
$\gamma=\frac{1}{\sqrt{1-v^{2}}}$ is the Lorentz factor, and we have set
$\lambda=1$ for simplicity. There is a topological charge $N$ associated
with a sine-Gordon kink
\begin{equation}
\label{N}
N=\frac{1}{2\pi}\int\limits_{-\infty}^{\infty}\frac{d\phi}{dx} {\rm d}x.
\end{equation}
At $x=\pm \infty$ the field $\phi$ needs to minimize the potential energy 
in (\ref{E}), hence we choose $\phi(-\infty) = 0$ and $\phi(\infty) = 2 
\pi N$. 
Therefore, equation (\ref{N}) clearly takes integer values.  For charge 
$N$ the Bogomolny bound becomes
\begin{equation}
E>\int\limits_{\phi(-\infty)=0}^{\phi(\infty)=2\pi
N}dW=8|N|\sqrt{\lambda}.
\end{equation}
For the case of multi-kinks this bound cannot be saturated as there is a
repulsive force between two static kinks.  Indeed, in
\cite{Perring:1962vs} Perring and Skyrme show that the asymptotic
interaction energy of two sine-Gordon kinks is given by
\begin{equation}
E_{int}=32 \sqrt{\lambda}e^{-\sqrt{\lambda}R},
\end{equation}
where $R$ is the separation between the kinks.

As well as being Lorentz invariant, the sine-Gordon model is also fully
integrable (see \cite{Scott} for a review of integrable systems).
Therefore, it has the feature that solution generating techniques, such as
the B\"acklund transformation, are applicable (see e.g.
\cite{Ablowitz:1991}). For $N=2$ this
transformation can be used to obtain the field
\begin{equation}
\label{2kinks}
\phi=4\arctan{\left(\frac{v\sinh{\left(\frac{x-vt}{1-v^2}\right)}}
{\cosh{\left(\frac{vt-v^2x}{1-v^2}\right)}}\right)}.
\end{equation}
This describes two individual kinks, one static at negative infinity in
time and the other travelling with speed $v$ at this instant. The field
configurations of the kinks gradually approach each other with time and 
the solitons interact. For simplicity, we have set $\lambda=1$ in
formula (\ref{2kinks}).

In the sine-Gordon model there is also a breather solution, which exists
in the charge zero sector and can be interpreted as a kink anti-kink
bound state. A solution that is a bound state of a breather and kink, the 
sine-Gordon wobble, is described in \cite{Kalbermann:2004fc}. These 
solutions show interesting behaviour when they are interacting with square 
wells \cite{Ferreira:2007ue, Ferreira:2007gu}.

\begin{figure}[!ht]
\begin{center}
\includegraphics[width=12cm]{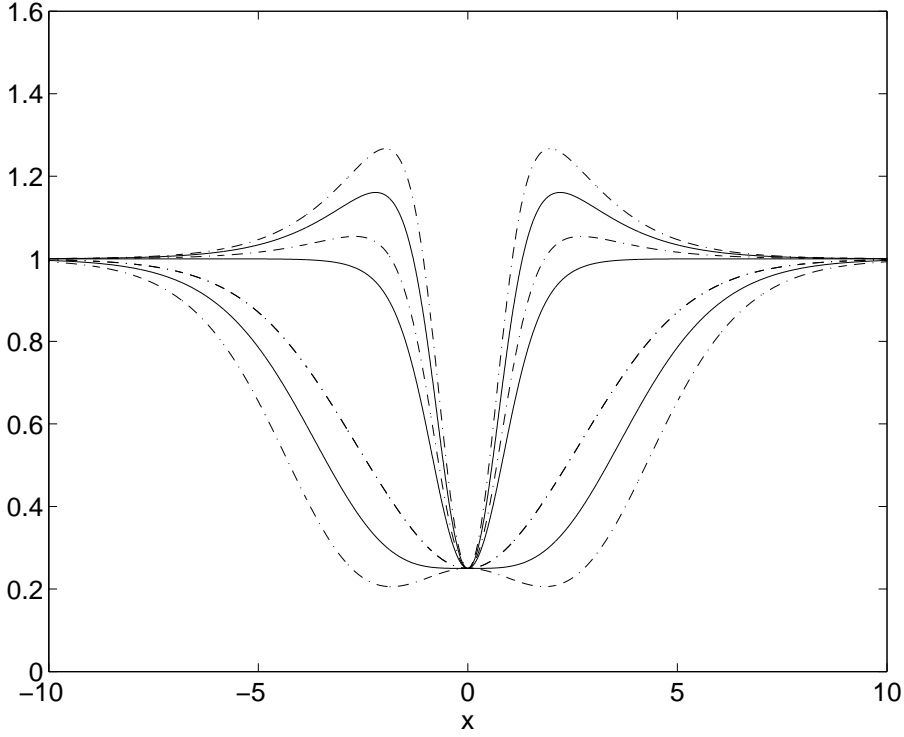} 
\caption{Wells with $\lambda(0)=\frac{1}{4}$ for various values of $a$ and
$b.$ Detailed scattering experiments have been performed on the wells
displayed with solids lines. These are the quartic well
$(a=\frac{5\sqrt{5}}{32}, b=\frac{3\sqrt{5}}{32}),$ a narrow well without
humps $(a=0,b=\sqrt{10}/4)$ and a generic well with humps $(a=-0.25,
b=1.126166088).$ Some other examples of wells are displayed
in dotted lines, for example the double well
$(a=\frac{3}{8},b=\frac{1}{8})$. \label{wells}} \end{center}
\end{figure}

\section{A class of space-dependent potentials}
\label{potentials}
In the following, we describe a class of space-dependent potentials, where 
the coupling constant $\lambda$ in (\ref{SGeq}) becomes space-dependent.
Square-well-type potentials have been discussed for example in
\cite{Piette:2006gw}.
These have the disadvantage that they are not smooth. Here, we propose a
smooth two-parameter family of barriers and wells, with the additional
advantage that there is an analytic solution for a static kink located
at the centre of the barrier or well. Figure \ref{wells} displays the
wells which we study in this paper, and Figure \ref{barriers} shows a
selection of barriers.

The static kink (\ref{sgkink}) located at $x=x_{0}$ solves the static field
equations
\begin{equation}
\label{kinkeq}
\partial_{xx} \phi - \lambda \sin(\phi) = 0,
\end{equation}
for constant $\lambda.$
By allowing $\lambda$ to depend on $x$ we can
find a two parameter family of kinks located at $x=0$ which solves the
static field equations (\ref{kinkeq}). The kinks are given by
\begin{equation}
\label{kinkatrest}
\phi(x) = \pi + 2\arctan\left(ax + b\sinh(x)\right)
\end{equation}
provided that we set
\begin{equation}
\lambda(x) = \frac{n_2 x^2 + n_1 x + n_0}{d_3 x^3 + d_2 x^2 + d_1 x +
d_0},
\end{equation}
where
\begin{eqnarray*}
n_2 &=&-a^2 b\sinh(x), \\
n_1 &=& 2 a\left(b^2+a^2+2ab\cosh(x)\right),\\
n_0 &=& b\sinh(x)\left(b^2+2a^2-1+4ab\cosh(x)+b^2\cosh(x)^2\right)
\end{eqnarray*}
and
\begin{eqnarray*}
d_3 &=& a^3,\\
d_2 &=& 3a^2b\sinh(x),\\
d_1 &=& a\left(1-3b^2+3b^2\cosh(x)^2\right),\\
d_0 &=& b\sinh(x)\left(1-b^2+b^2\cosh(x)^2\right).
\end{eqnarray*}
Note that the trivial vacuum solutions $\phi(x) \equiv 2 \pi n$ also 
satisfy (\ref{kinkeq}) for all functions $\lambda(x)$ provided $n$ is an 
integer.

It is easy to see that $\lambda(x) = \lambda(-x)$ and that the map
\begin{equation}
(a,b) \mapsto (-a,-b)
\end{equation}
leaves $\lambda(x)$ invariant. Physically, this means that $\lambda$ is
the same for a kink $(b>0)$ and an anti-kink $(b<0)$. In the following, we
restrict our attention to kinks and set $b \ge 0$. Since the $\sinh(x)$
term dominates for large $|x|$ the parameter $a$ can take positive and
negative values provided $b > 0.$ Setting $b>0$ also guarantees that
\begin{equation}
\lim\limits_{|x| \to \infty} \lambda(x) = 1,
\end{equation}
which means that the kink located at $x_0$ given by (\ref{sgkink})
tends to the exact solution for large $|x_0|,$ that is when the
kink is far enough away from the well. For $b=0$, the
asymptotics change. In this case,
\begin{equation}
\lambda(x) =  \frac{2a^2}{a^2x^2+1},
\end{equation}
which tends to $0$ as $|x| \to \infty.$ Since we want the usual
sine-Gordon kink (\ref{sgkink}) to be an asymptotic solution, we only
consider the case $b>0.$

In order to determine the values of $a$ and $b$ for which $\lambda(x)$ is
regular, we solve equation (\ref{kinkeq}) for $\lambda$ and obtain
\begin{equation}
\lambda(x) = \frac{\partial_{xx} \phi}{\sin(\phi)}.
\end{equation}
Therefore, singularities can only occur for $\sin \phi = 0$. With our
choice of $b>0$ we have
\begin{equation}
-\frac{\pi}{2} < \arctan\left(ax + b\sinh(x)\right) < \frac{\pi}{2},
\end{equation}
hence, since $\arctan$ is a monotonic function and $\arctan(0) = 0$,
$\lambda(x)$ can only have singularities at $g(x) = 0$, where
\begin{equation}
\label{ax+bsinhx}
g(x) = ax + b \sinh(x).
\end{equation}
Taking the derivative of (\ref{ax+bsinhx}) leads to
\begin{equation}
g^\prime(x)  = a + b\cosh(x).
\end{equation}
This implies that $g^\prime(x) > 0$ for $a + b > 0,$ so that $g(x)$ is a
monotonic function whose only root is $x=0$. We can evaluate the limit
\begin{equation}
\label{lim1}
\lim\limits_{x \to 0} \lambda(x) = \frac{2(a+b)^3-b}{a+b}.
\end{equation}
Hence, $\lambda(x)$ diverges for $b>0$ and $a \to -b.$ For $a+b < 0,$ the
function $g(x)$ has one maximum and one minimum and hence two non-trivial
zeros. Generically, $\lambda(x)$ will be singular for $a + b \le 0.$

Physical applications usually demand that $\lambda(x) > 0.$
This condition is satisfied provided $a$ satisfies the stronger inequality
\begin{equation}
a + b> \left(\frac{b}{2}\right)^\frac{1}{3}.
\end{equation}
Furthermore, we are mostly interested in wells located at the origin, 
which implies $\lambda(0)<1.$

An interesting one parameter family is obtained by setting $a=0.$ It
follows from  (\ref{lim1}) that $\lambda(0)= 2b^2-1.$ Therefore, we
obtain a well with $\lambda(x)>0,$ for $\sqrt{\frac{1}{2}}<b<1,$ whereas
for $b>1$ this is a barrier. This one-parameter family is an example of a
pure well or barrier, respectively, because the only extrema of
$\lambda(x)$ is at $x=0.$

Fixing the value of $\lambda(0)=h$ and $b$ gives a cubic equation for
$a,$ which is most conveniently expressed in terms of the variable $a+b,$ 
namely
\begin{equation}
\label{cubic}
(a+b)^3 - \frac{h}{2} (a+b) -\frac{b}{2} = 0,
\end{equation}
This cubic has the discriminant
\begin{equation}
D = -\frac{h^3}{216} + \frac{b^2}{16}.
\end{equation}
Hence equation (\ref{cubic}) has three real solutions for $D<0,$ two real
solutions for $D=0$ and only one real solution for $D>0$. However, not all
solutions will lead to regular wells.

In order to gain a better understanding of the various wells and barriers
it is useful to calculate
\begin{equation}
\label{lim2}
\lim\limits_{x \to 0} \frac{d^2\lambda(x)}{dx^2} =
\frac{-12(a+b)^6+12b(a+b)^3-b(a+b)+b^2}{2(a+b)^2}.
\end{equation}
When this limit vanishes we have a quartic well.

In Figure \ref{wells}, we considered wells with $\lambda(0)=\frac{1}{4}.$
There is a one-parameter family of wells with humps on both sides for
$a<0.$ Our main example is $a=-0.25$ and $b= 1.126166088.$ There
is another one parameter family of wells with $a>0$ which includes pure
wells, the quartic well and double wells. As can be seen in Figure
\ref{wells}, the pure wells for $a>0$ are wider than the pure well at
$a=0.$ Using equation (\ref{lim2}) we can derive the values for the
quartic well, namely, $a=\frac{5 \sqrt{5}}{32}$ and $b =
\frac{3\sqrt{5}}{32}$ with $\lambda(0) = \frac{1}{4}.$ For
$a>\frac{5 \sqrt{5}}{32}$ we obtain double wells, where there is a local
maximum at $x=0,$ and global minima at either side.

In Figure \ref{barriers}, we show interesting examples of barriers. Again,
there are pure barriers, barriers with wells at
either side and volcano (or double peak) barriers.

\section{Dynamics of one kink in the presence of wells and barriers}
\label{onekink}
In this section, we consider a sine-Gordon kink travelling towards our
standard well with various initial velocities and plot the trajectories. 
We also calculate the critical velocities for different barriers and 
compare these to an analytic approximation. First, we briefly comment on 
our numerical scheme.

The equations of motion for the kinks have been solved using a standard
fourth order Runge-Kutta method with gridsize 10001. Plus and minus
``infinity'' are located at $50$ and $-50,$ respectively, so that the
stepsize in space is $\Delta x = 0.01.$ The stepsize in time has
usually been taken to be $\Delta t = 0.0001$ which is the same choice of
parameters as in \cite{Piette:2006gw}. For small initial velocities of the
kink, larger values of $\Delta t$ are appropriate.

\begin{figure}[!ht]
\begin{center}
\subfigure[$v=0.2$\label{1kinka}]{
\includegraphics[width=7cm]{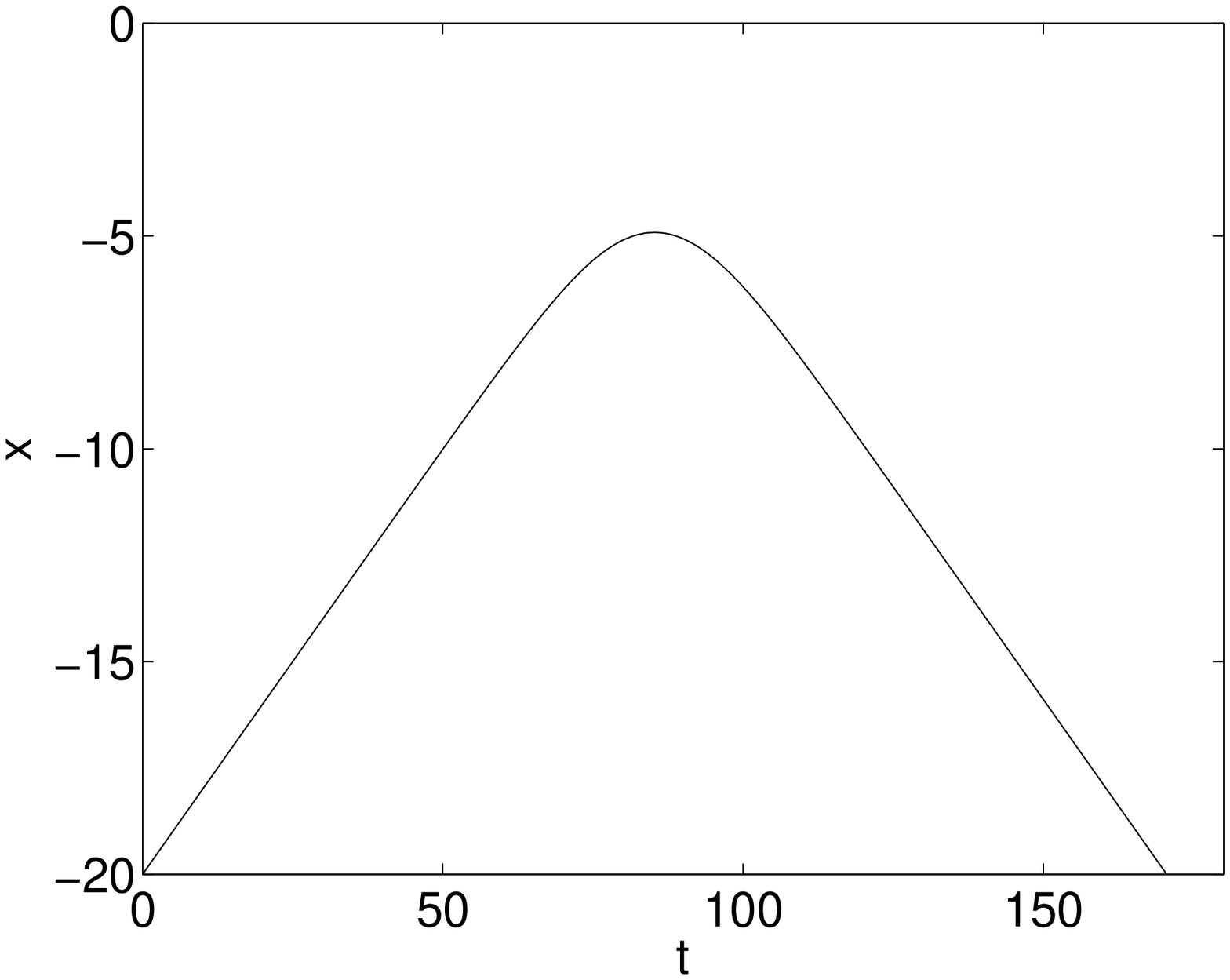}
} 
\quad
\subfigure[$v=0.33$\label{1kinkb}]{
\includegraphics[width=7cm]{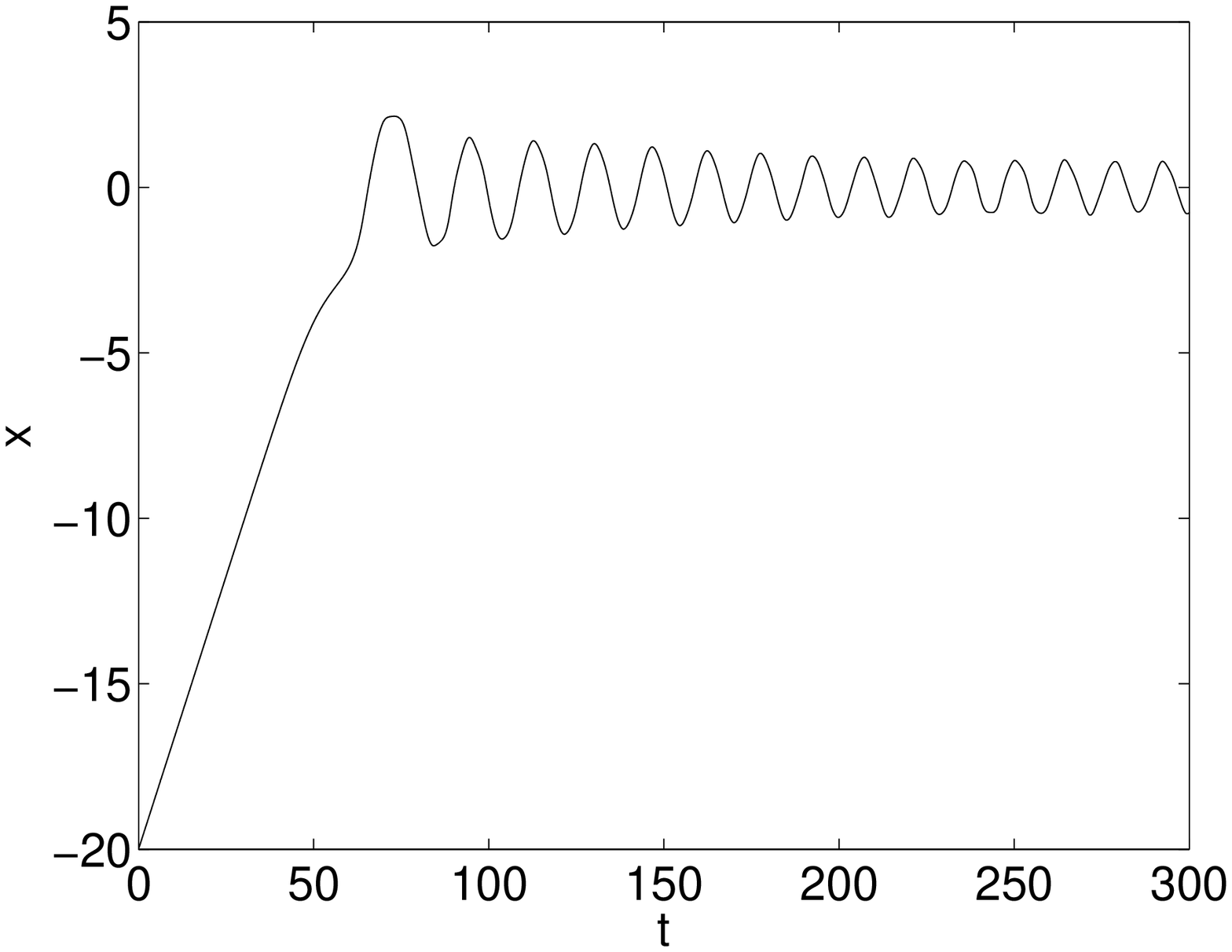}}
\quad
\subfigure[$v=0.3873$\label{1kinkc}]{
\includegraphics[width=7cm]{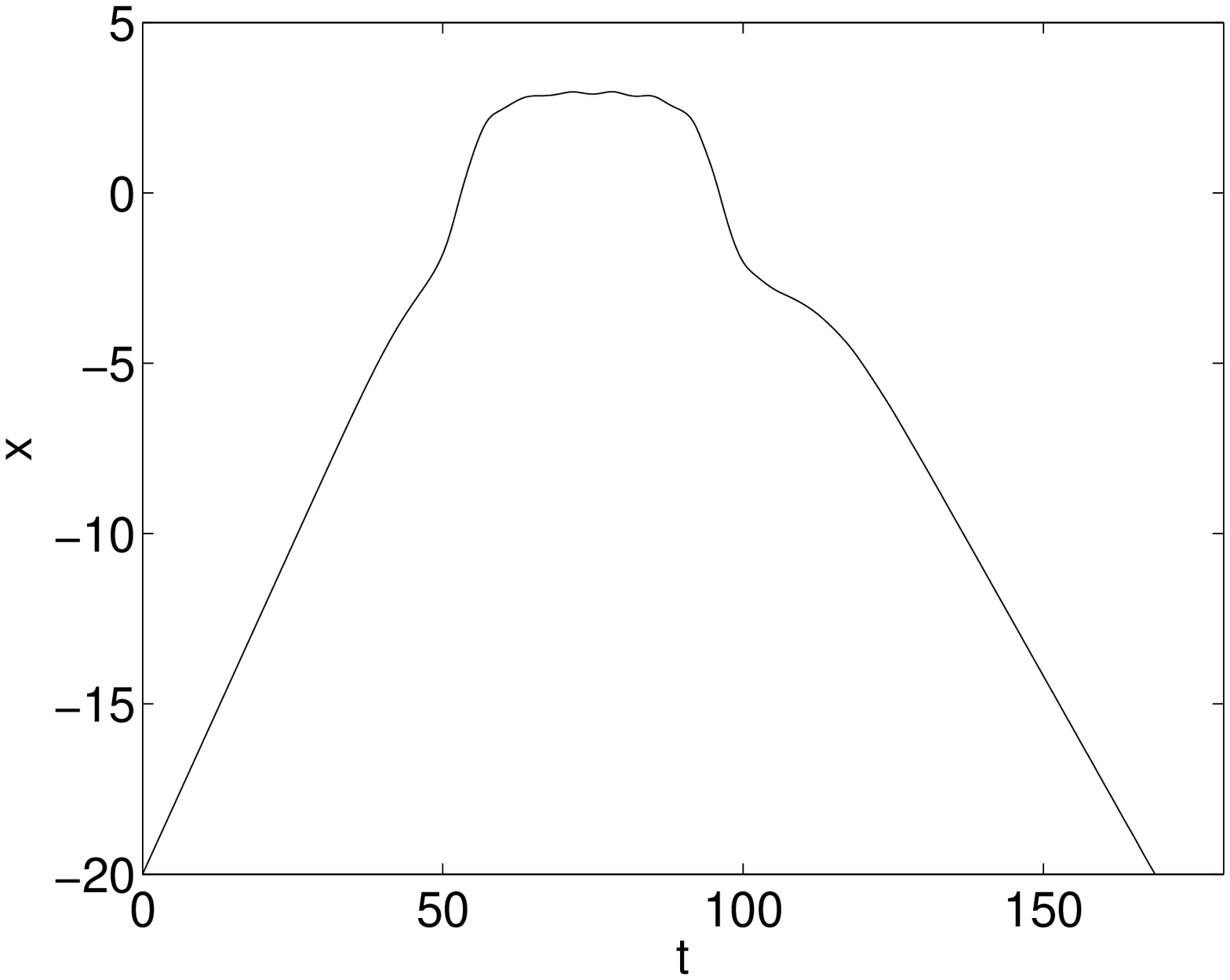}}
\quad
\subfigure[$v= 0.5$\label{1kinkd}]{
\includegraphics[width=7cm]{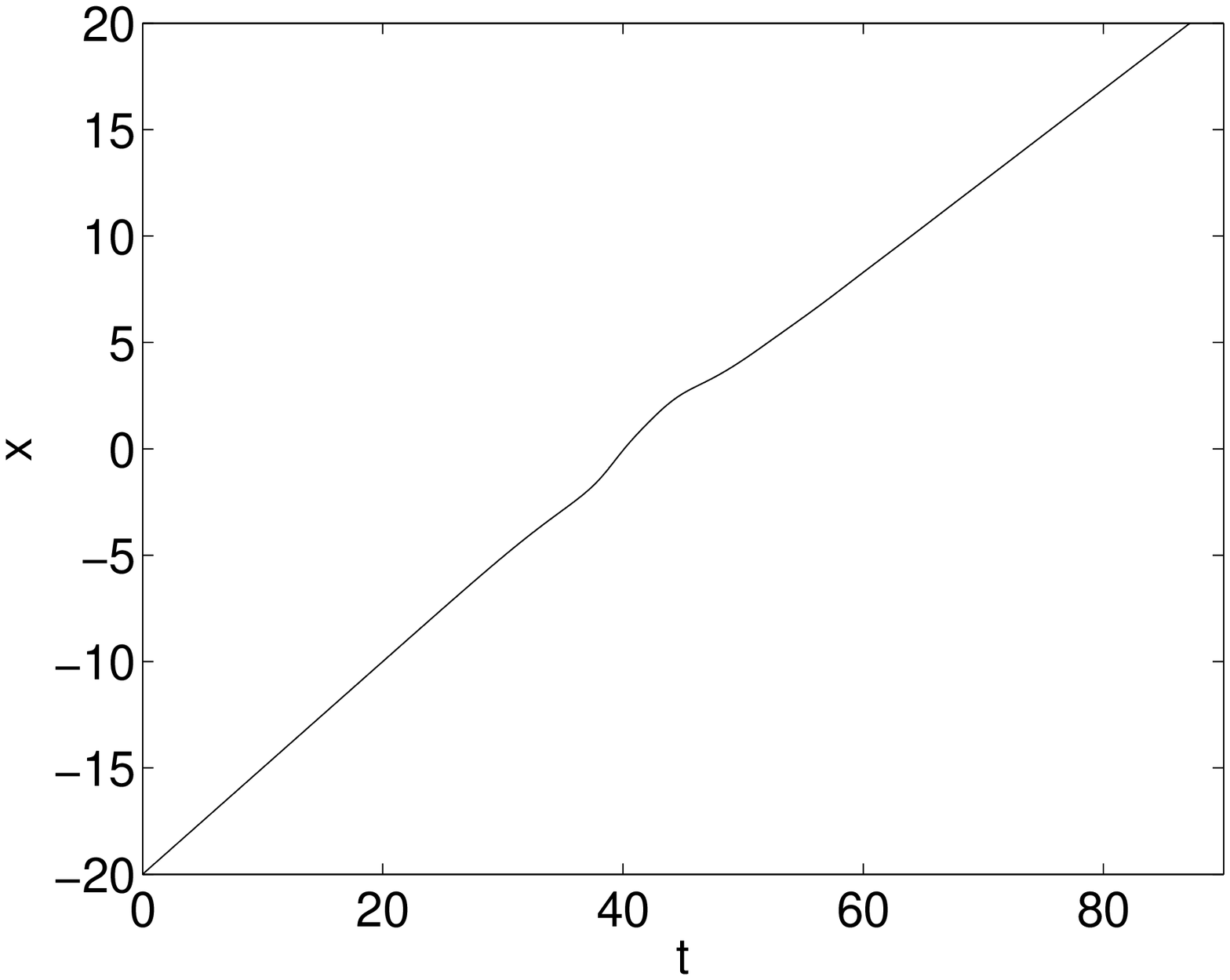}}
\caption{Kink trajectories $x(t)$ for a well with $a=-0.25$ and
$b=1.126166088.$
Figure \ref{1kinka} shows elastic back-reflection.
In Figure \ref{1kinkb} the kink becomes trapped in the well. In Figure
\ref{1kinkc} the
kink is back-reflected by the well. In Figure \ref{1kinkd} the kink crosses
the well.
\label{1kink}} \end{center}
\end{figure}

\subsection{Plots of trajectories}
Figure \ref{1kink} illustrates all relevant phenomena for one kink
interacting with a well. We have chosen the well with parameters
$a=-0.25$ and $b=1.126166088$, which has a minimum
$\lambda(0)=\frac{1}{4}$ and two humps at $x\approx \pm 2.20$ of height
$1.16$.

For low velocity $v < 0.3$, there is elastic back-reflection from the
hump, so that initial and final velocity are equal in size, as in Figure
\ref{1kinka}. For higher velocity the kink overcomes the initial barrier
and becomes trapped in the well which is illustrated in Figure
\ref{1kinkb}.
The critical velocity $u_c$ is defined as the smallest initial velocity
which allows the kink to cross the well (or barrier). For the well
discussed here, the critical velocity is $u_c \approx 0.387.$ An example
of a kink crossing the well is given in Figure \ref{1kinkd} where it can
be seen that the kink gains speed inside the well. The most interesting
behaviour happens for a narrow range of velocities just below $u_c.$ As
Figure \ref{1kinkc} shows, the kink may enter the well, get reflected from the
well and then leave the well travelling in the opposite direction. This
type of behaviour has been discussed in detail in \cite{Piette:2006gw}.
For our standard well, this resonance window is very narrow. In fact the
kink is trapped for $v =0.38735,$ is back-reflected for $v=0.38736$ and
trapped once more for $v=0.38737.$ For $v=0.38738$ the kink escapes.

\subsection{Critical velocities for one kink interacting with a barrier}
\label{criticalvelocity}
In this section we follow an argument in \cite {AlAlawi:2009rt} to
calculate an analytic approximation to the critical velocity, $u_c$.
We then compare the analytic calculations with numerical results.

Making use of the moduli space approximation \cite{Manton:1981mp}, we make
the ansatz
\begin{equation}
\label{phiX(t)}
\phi(x;X)=4\arctan\left(\exp\left(x-X(t)\right)\right),
\end{equation}
where $X(t)$ is the position of the solution as a function of time.
Substituting (\ref{phiX(t)}) into (\ref{lagrange}) we obtain
\begin{equation}
{\cal L}=\frac{8{\dot
X}^{2}e^{2(x-X)}}{(1+e^{2(x-X)})^2}-\left(1+\lambda(x)\right)
\frac{8e^{2(x-X)}}{(1+e^{2(x-X)})^2}.
\end{equation}
Note that $\lambda$ is now a function of $x.$ The Lagrangian is then
\begin{equation}
L=4({\dot X}^{2}-1)-\int\limits_{-\infty}^{\infty}
\frac{8 \lambda(x) e^{2(x-X)}}{(1+e^{2(x-X)})^2}\,{\rm d}x.
\end{equation}
This has a non-trivial potential term, and the motion of a soliton on the
moduli space is therefore not geodesic for non-constant $\lambda(x).$ This
is expected as there will always be a static force on the solution when in
the presence of a well or barrier such as the one in the model
presented.

\begin{figure}[!htb]
\begin{center}
\includegraphics[width=12cm]{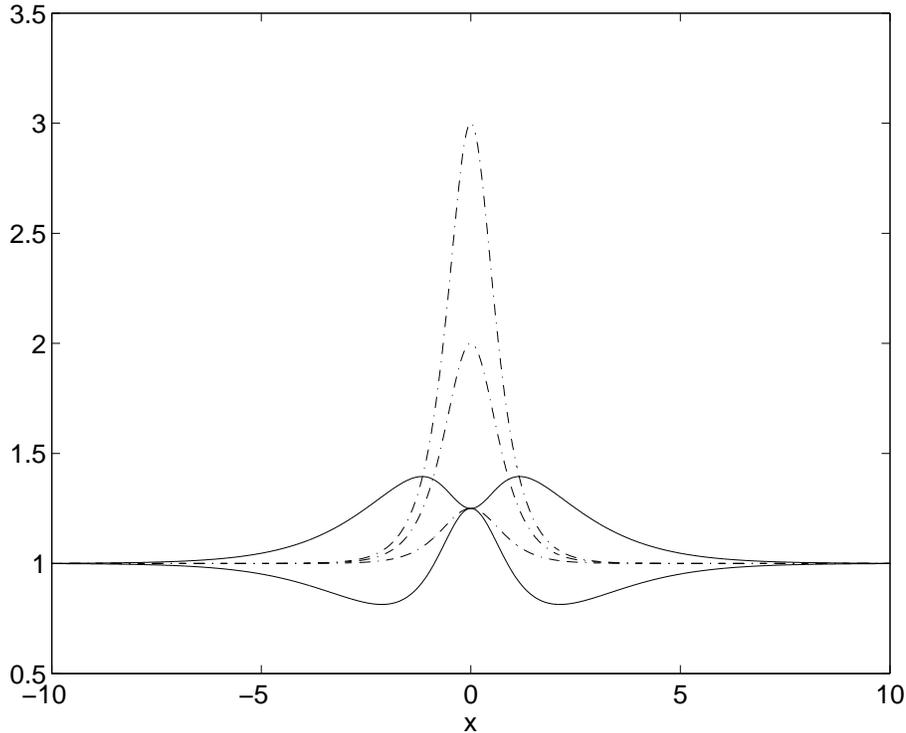} 
\caption{Different types of barriers:
a barrier with two side wells ($a=0.25,$ $b=0.75$), a volcano barrier 
($a=-0.5,$ $b=1.657970214$), and three pure barriers
($a=0,$ $b=\frac{3\sqrt{2}}{4}$), ($a=0,$
$b=\frac{\sqrt{6}}{2}$) and ($a=0,$ $b=\sqrt{2}$) with height $1.25,$ $2$ 
and $3,$ respectively.
\label{barriers}}
\end{center}
\end{figure}

The total energy of a soliton can be calculated
\begin{equation}
\label {int}
E  =  \frac{1}{2}8{\dot
X}^2+4+\int\limits_{-\infty}^{\infty}
\frac{8 \lambda(x) e^{2(x-X)}}{(1+e^{2(x-X)})^2}\,
{\rm d}x.
\end{equation}
In the absence of an obstacle, $\lambda(x) \equiv 1,$ and the energy is
\begin{equation}
E=\frac{1}{2}8{\dot X}^2+8.
\end{equation}
For a soliton at rest i.e. $\dot X=0$ the kink mass is
\begin{equation}
M_{rest}=8.
\end{equation}
In the case of a pure barrier, $\lambda(x)>1,$ therefore, for a kink 
stationary on top of the barrier
\begin{equation}
M_{rest}>8.
\end{equation}
For a pure well, on the other hand, $\lambda(x)<1,$ so a static kink at 
the bottom of the well will have mass
\begin{equation}
M_{rest}<8.
\end{equation}
Otherwise, one cannot say which inequality holds without explicitly
evaluating the integral appearing in (\ref {int}).
Energy conservation dictates that in the case of elastic scattering
the mass, $M_{B},$ of a static kink
on top of a barrier, is related to the critical velocity,
$u_{c},$ for the barrier by the following formula (appearing in
 \cite{AlAlawi:2009nr})
\begin{equation}
\label{crit}
\frac{8}{\sqrt{1-{u_{c}}^2}}=M_{B}.
\end{equation}
Figure \ref{barriers} shows some barriers of different types and
heights.

In table \ref{criticalv} we present critical velocities calculated both
numerically and using equation ($\ref {crit}$).  It is clear that the 
theoretical approach is
accurate for barriers that are monotonically increasing before the origin
and monotonically decreasing afterwards.  Otherwise, there is discrepancy.
This is explained by considering the fact that for all models investigated
so far solitons have elastic scattering on barriers and inelastic
scattering on holes.

\begin{table}[!ht]
\begin{center}
\begin{tabular}{|l|l|l|l|l|l|l|}\hline
& & & & & & \\
$a$ & $b$ & mass & $u_c$ (theoretical)&$u_c$ (numerical)&
$\lambda(0)$ &
Type of
barrier\\
& & & & & & \\
\hline
0.25&  0.75 & 8.33 & 0.271 & 0.280& 1.25 &
Barrier with wells \\
\hline
-0.50 &1.66& 9.17&  0.489 & 0.505 & 1.25 & Volcano \\
\hline
0.00& 1.06& 8.64 & 0.379 & 0.379 & 1.25 & Pure
barrier \\
\hline
0.00& 1.22 &10.35&0.635 &0.636 & 2.00 & Pure barrier\\
\hline
0.00& 1.41&12.28&0.759 &0.762 & 3.00 & Pure barrier\\
\hline
\end{tabular}
\caption{The values of $a$ and $b$ are the parameters of the barriers
discussed in Figure \ref{barriers}, the column ``mass'' gives the energy
of the a kink located on top of the barrier, the two columns with
$u_c$ compare the analytic approximation to the critical velocity to
its numerical value, $\lambda(0)$ is the height of the barrier at $x=0$
and the final column gives the type of barrier.\label{criticalv}}
\end{center}
\end{table}

\begin{figure}[!ht]
\begin{center}
\subfigure[$v=0.2$\label{2kinka}]{
\includegraphics[width=7cm]{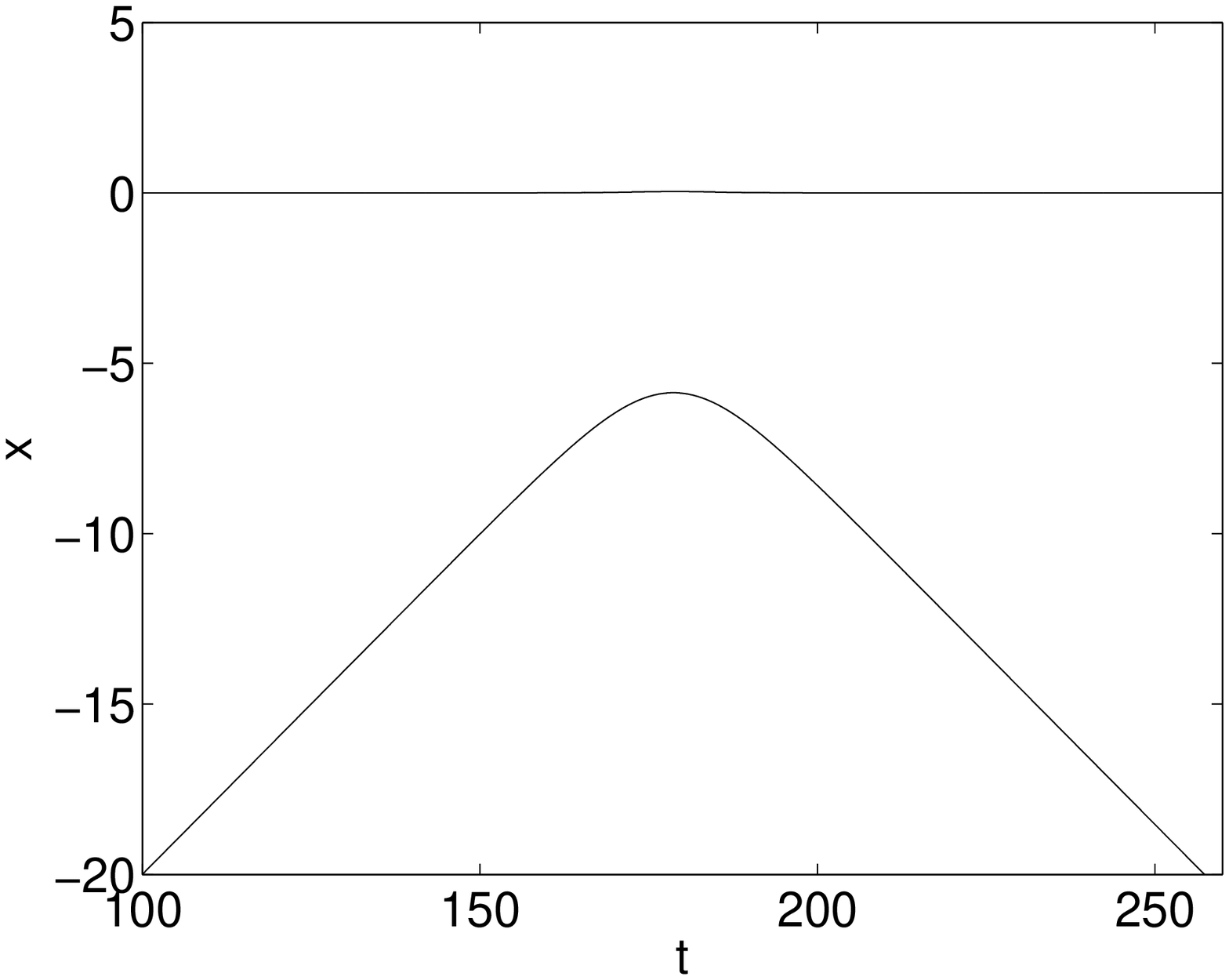}
} 
\quad
\subfigure[$v=0.6$\label{2kinkb}]{
\includegraphics[width=7cm]{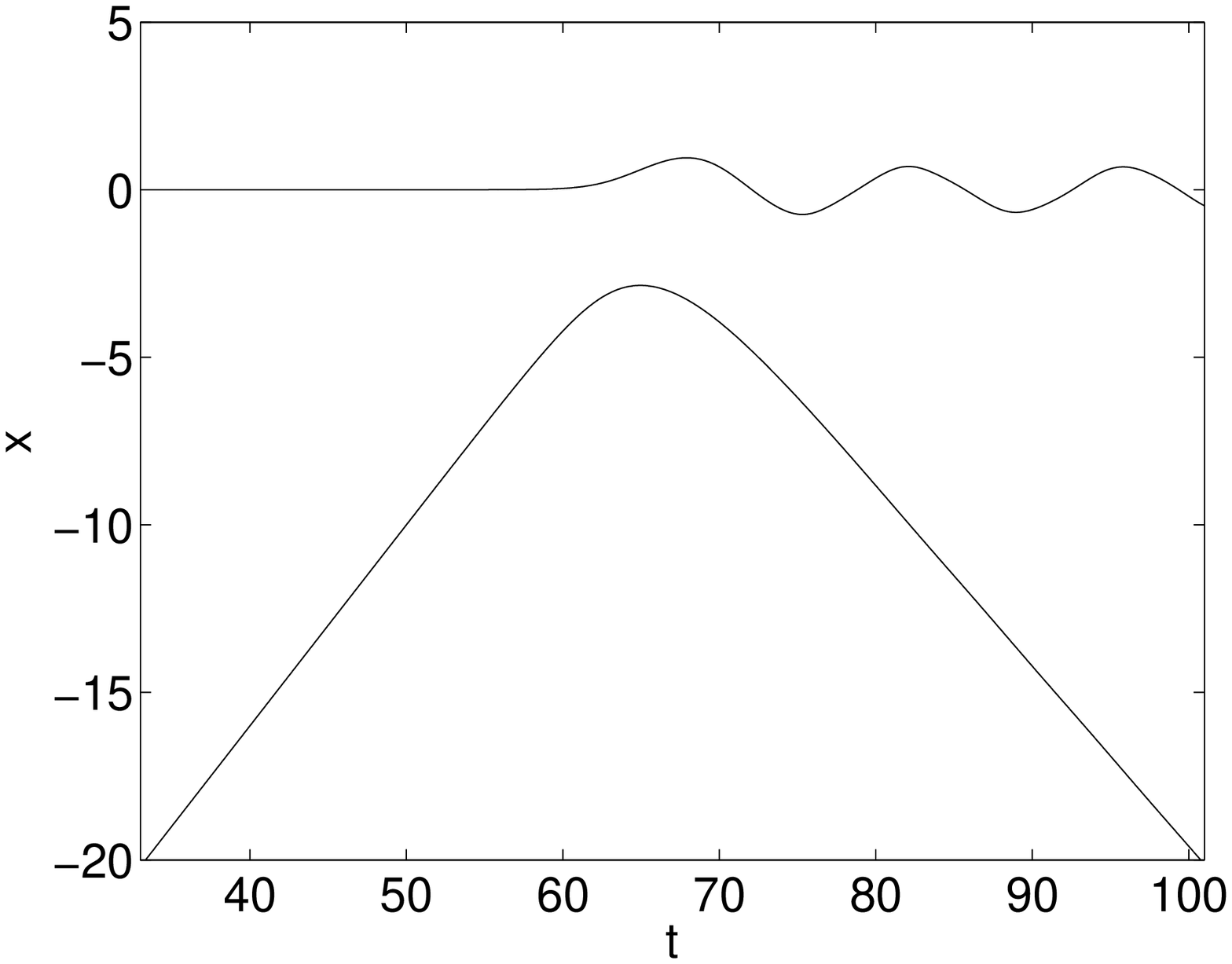}}
\quad
\subfigure[$v=0.7$\label{2kinkc}]{
\includegraphics[width=7cm]{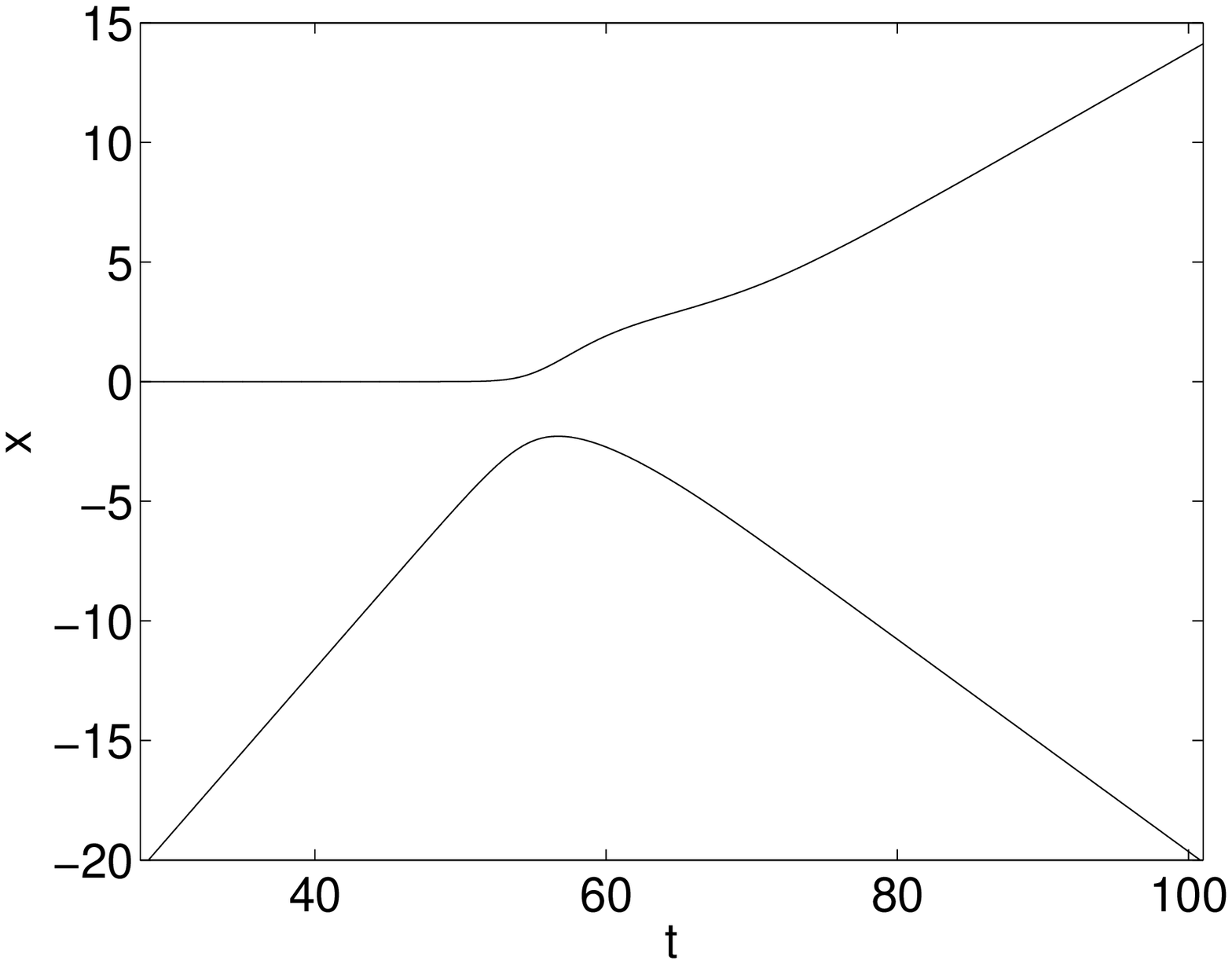}}
\quad
\subfigure[$v= 0.75$\label{2kinkd}]{
\includegraphics[width=7cm]{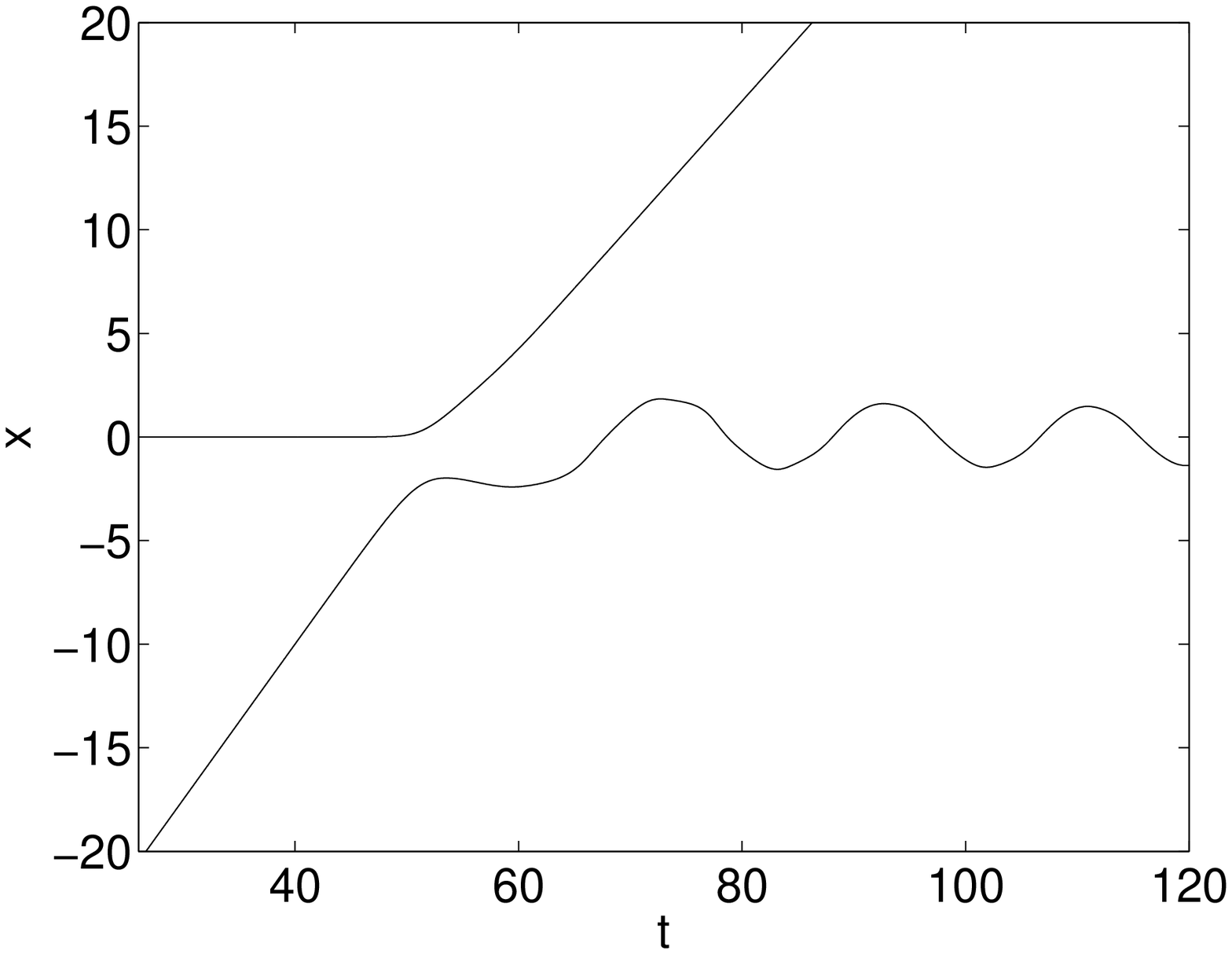}}
\caption{Kink trajectories $x(t)$ for a well with $a=-0.25$ and
$b=1.126166088.$ One kink is travelling towards the well whereas a
second kink is stationary inside the well.
Figure \ref{2kinka} shows elastic back-reflection, the
kink in the well remains stationary. In
Figure \ref{2kinkb} the kink experiences inelastic scattering, the trapped
kink oscillates in the well. In Figure
\ref{2kinkc} the first kink is back-reflected while the second kink
escapes from the well. In Figure \ref{2kinkd} the first kink becomes
trapped while the second kink escapes from the well.\label{2kink}}
\end{center}
\end{figure}

\section{Dynamics of two kinks in the presence of a well}
\label{twokinks}
In general, the interaction of two kinks with a well is rather
complicated because the trajectories depend on the initial positions and
the initial velocities of both kinks. Here, we restrict our attention to
scattering processes when the first kink approaches the well with given
initial velocity $v,$ and the second kink is at rest in the well. Our
potential $\lambda(x)$ has been chosen, such that the first kink is given
asymptotically by equation (\ref{initialkink}) whereas the second kink
is given by equation (\ref{kinkatrest}). The exact solution
(\ref{2kinks}) cannot be generalized to our wells. However, since
kinks are exponentially localized, we can just concatenate the two kinks
provided the first kink is far enough away from the well.

Figure \ref{2kink} illustrates some important trajectories for our
standard well. When the first
kink travels with small initial velocity, e.g. $v=0.2,$ then there is
elastic back-reflection, such that the final velocity of the first kink is
$-v,$ and the second kink remains at rest in the well, see Figure
\ref{2kinka}. The trajectory of the first kink closely resembles the
trajectory of a single kink in Figure \ref{1kinka}. Figure \ref{2kinkb}
shows inelastic scattering where the first kink loses energy, and the
second kink clearly gains energy because is it now oscillating in the
well. In Figure \ref{2kinkc} the second kink is knocked out of the well by
the first kink which also escapes from the well. Figure \ref{2kinkd} shows
a similar scenario. The second kink is again knocked out of the well,
whereas the first kink is now trapped in the well. It is worth comparing
this outcome to the exact solution when there is no well. In this case,
the first kink loses all its kinetic energy to the second kink, so that
after the scattering event the first kink is at rest whereas the second
kink moves with velocity $v.$

We checked the range  $0.695 < v < 0.6951$ in detail, in the hope that
there would be a double back-reflection, so that first kink and second kink
both escape on the same side. But for our standard well, we were unable to
find this kind of behaviour.

\begin{figure}[!ht]
\begin{center}
\subfigure[$v=0.01$\label{2kinke}]{
\includegraphics[width=7cm]{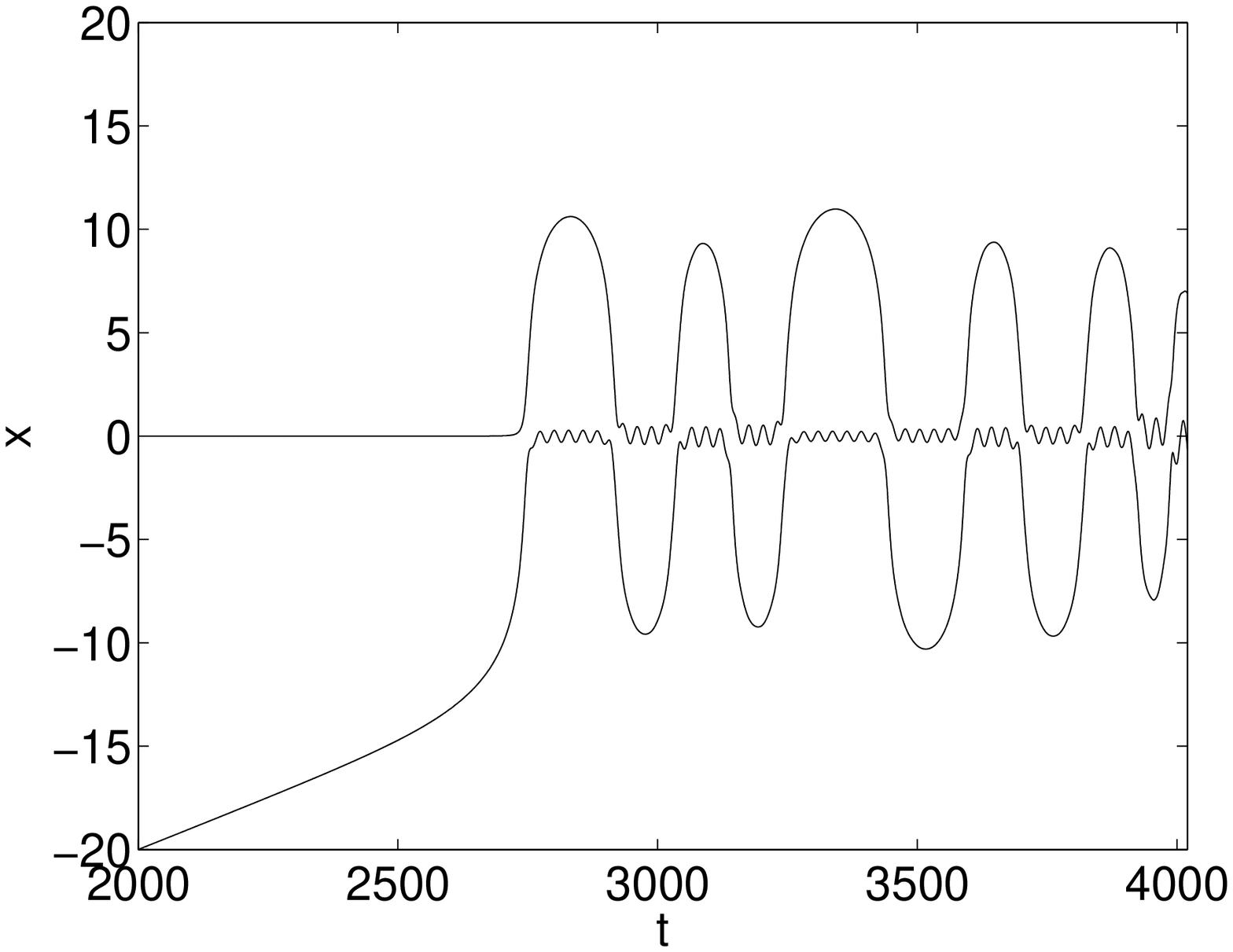}}
\quad
\subfigure[$v=0.025$\label{2kinkf}]{
\includegraphics[width=7cm]{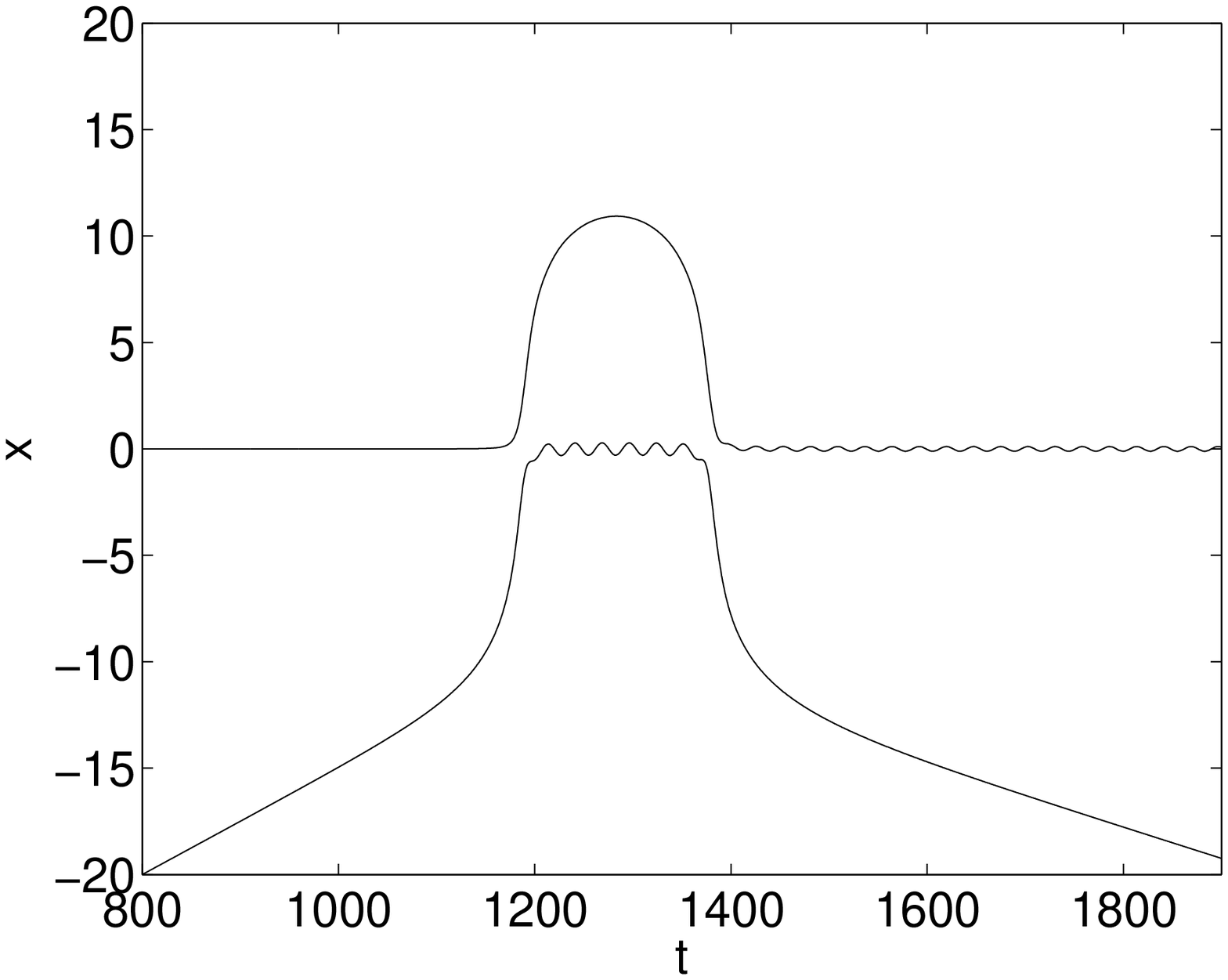}}
\caption{Kink trajectories $x(t)$ for the quartic well ($a=\frac{5
\sqrt{5}}{32},$  $b = \frac{3\sqrt{5}}{32}$).
One kink is travelling towards the well with velocity $v$ whereas a
second kink is stationary inside the well.
In Figure \ref{2kinke} both kinks become trapped in the well.
Figure \ref{2kinkf} shows a novel type of back-reflection.\label{2kinkq}}
\end{center}
\end{figure}

In Figure \ref{2kinkq}, we discuss two interesting trajectories for the
quartic well ($a=\frac{5 \sqrt{5}}{32},$  $b = \frac{3\sqrt{5}}{32}$).
This is a rather wide well which allows both kinks to be trapped. Figure
\ref{2kinke} shows an interesting quasi-periodic motion where the two
kinks knock each other out of the centre of the well but not gaining
enough energy to leave the well. Figure \ref{2kinkf} shows a novel type of
back-reflection. The first kink enters the well and speeds up, then hits
the second kink and knocks it off the centre of the well. The second kink
slows down, returns to the centre and gives the first kink enough kinetic
energy to escape the well.

\begin{figure}[!ht]
\begin{center}
\subfigure{
\includegraphics[width=7cm]{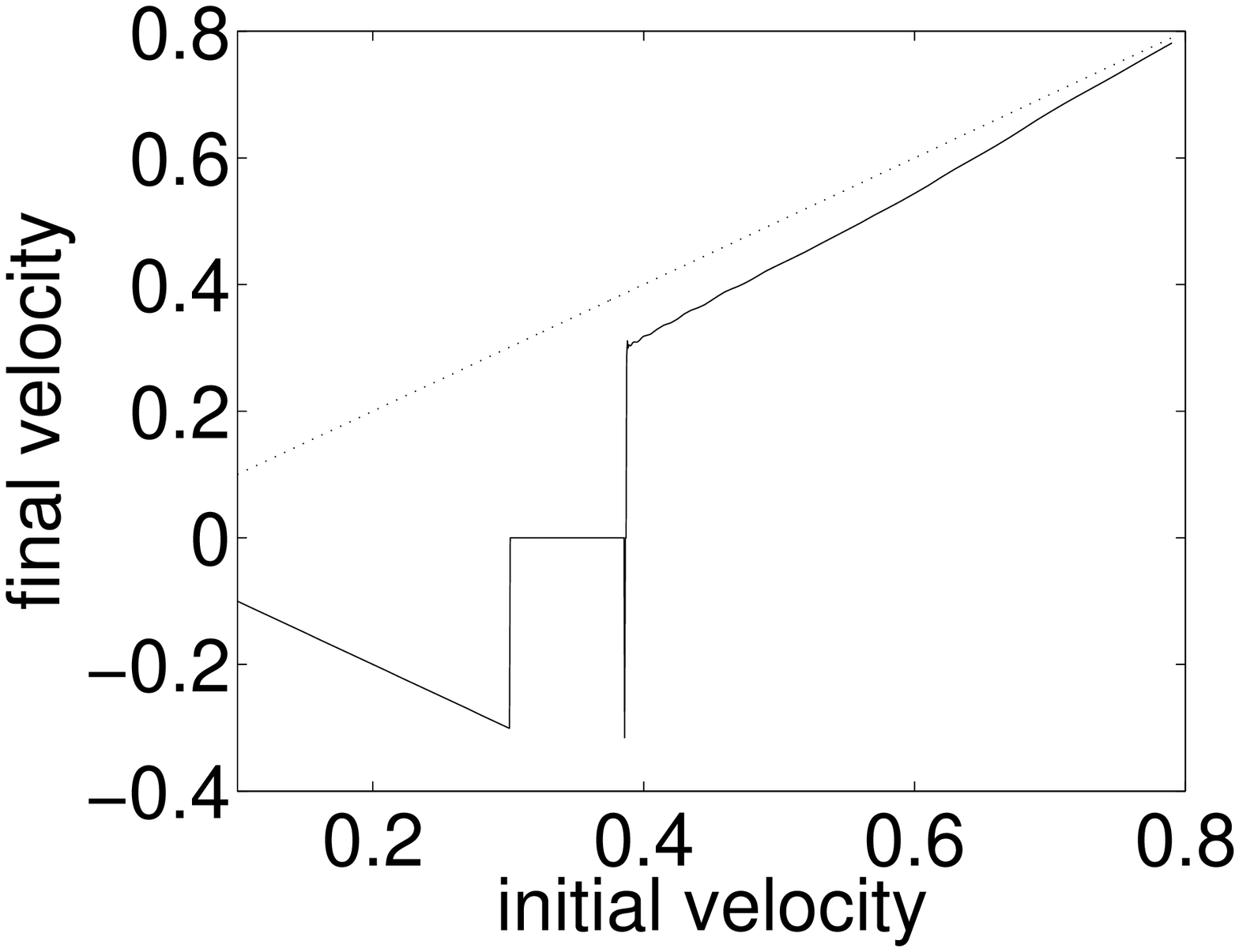}}
\quad
\subfigure{
\includegraphics[width=7cm]{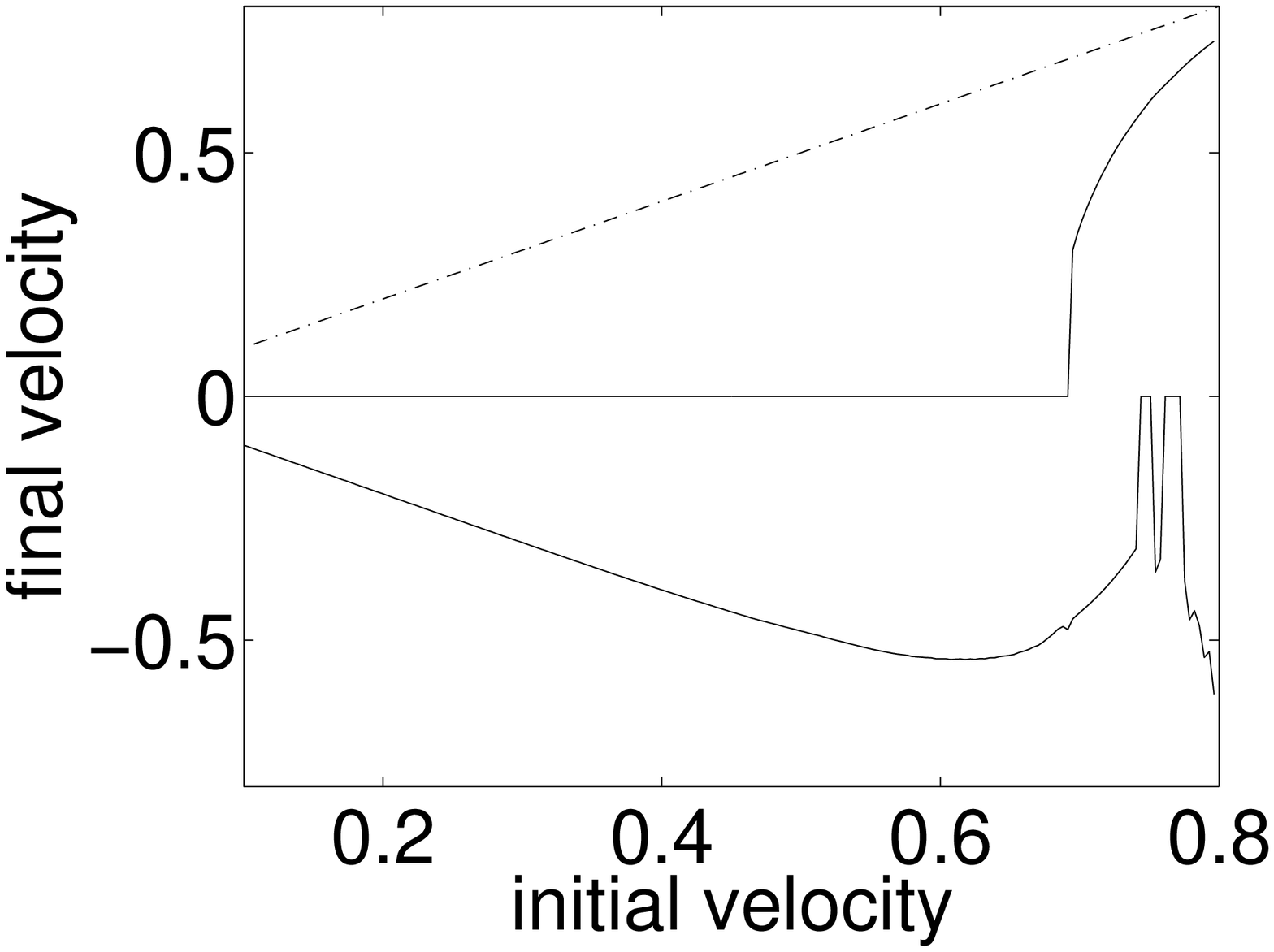}}
\caption{Initial velocity versus final velocity for a kink incident on
a well with humps with $a=-0.25$ and $b=1.126166088.$ For the figure on
the left, the well is empty whereas for the figure on the right, there is
a kink in the well, which is initially at rest.\label{vivf1}}
\end{center}
\end{figure}

Figure \ref{vivf1} shows initial versus final velocity for our standard
well. In the left figure, one kink travels towards the well with initial
velocity $v.$ For $v<0.3$ the kink is elastically back-reflected. As $v$ 
is increased, the kink is trapped. 
For $v>0.387$ the kink escapes the well. As 
$v$ tends to one the kink feels the influence of the well less, so that 
the final velocity tends to the initial velocity for $v \approx 1$. The
back-reflection at $v=0.38736$ is too narrow to be detected with the 
resolution used for this figure. However, there is another backreflection 
at $v=0.3860.$

The right figure in \ref{vivf1} shows initial versus final velocity for
our standard well, where a second kink is at rest inside the well. Elastic
scattering takes place for a longer range of initial velocity $v.$ As the 
initial velocity of the incoming kink increases, the kink in the well is 
set oscillating more energetically.  Because of this, at some point the 
final velocity of the first kink starts to become less negative.  At 
higher initial velocities the final velocity of the initially static kink 
increases and approaches one, while the final velocity of the other 
soliton takes the value zero in two separated ``windows'', but is more 
negative between and after these windows.       

\begin{figure}[!ht]
\begin{center}
\subfigure{
\includegraphics[width=7cm]{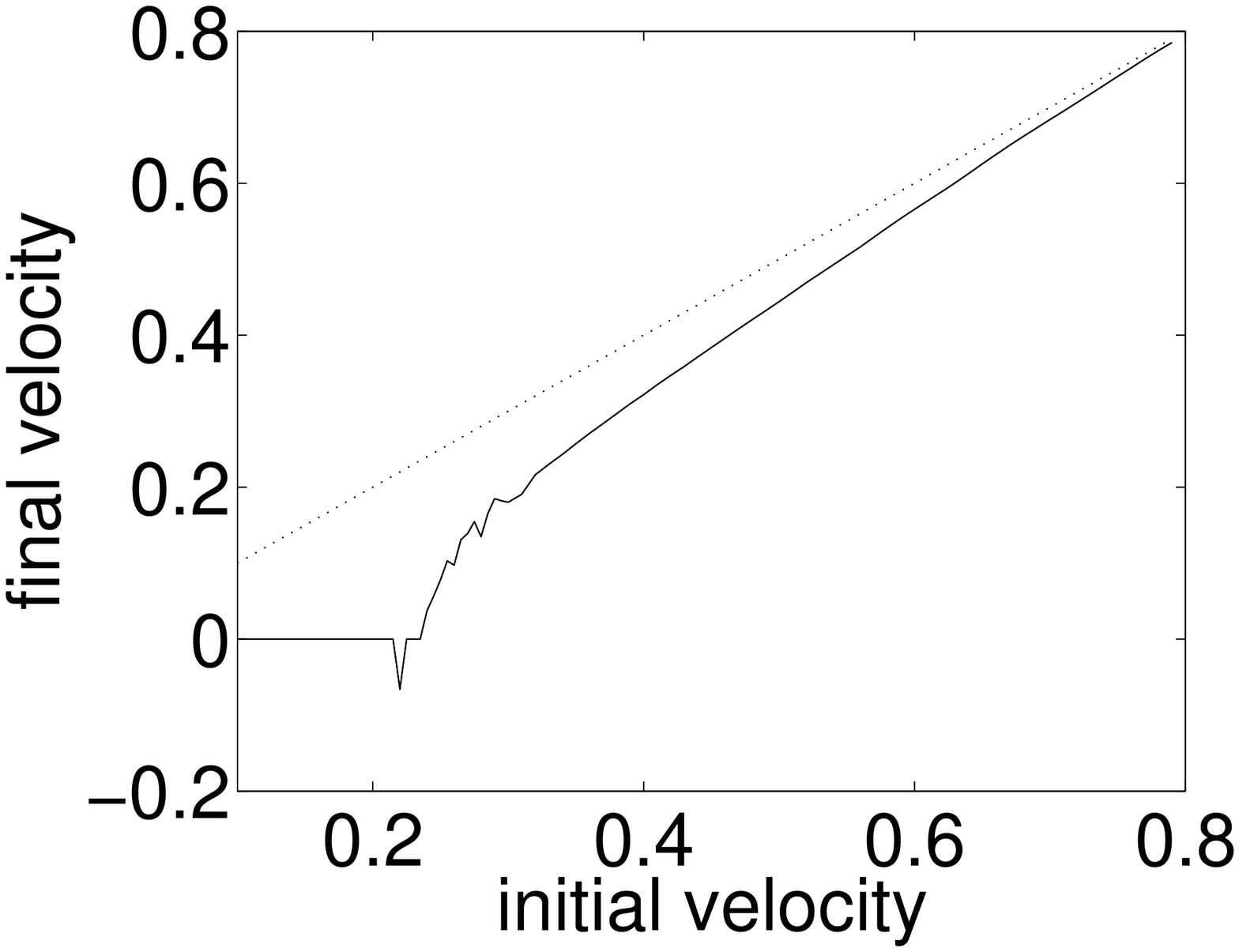}}
\quad
\subfigure{
\includegraphics[width=7cm]{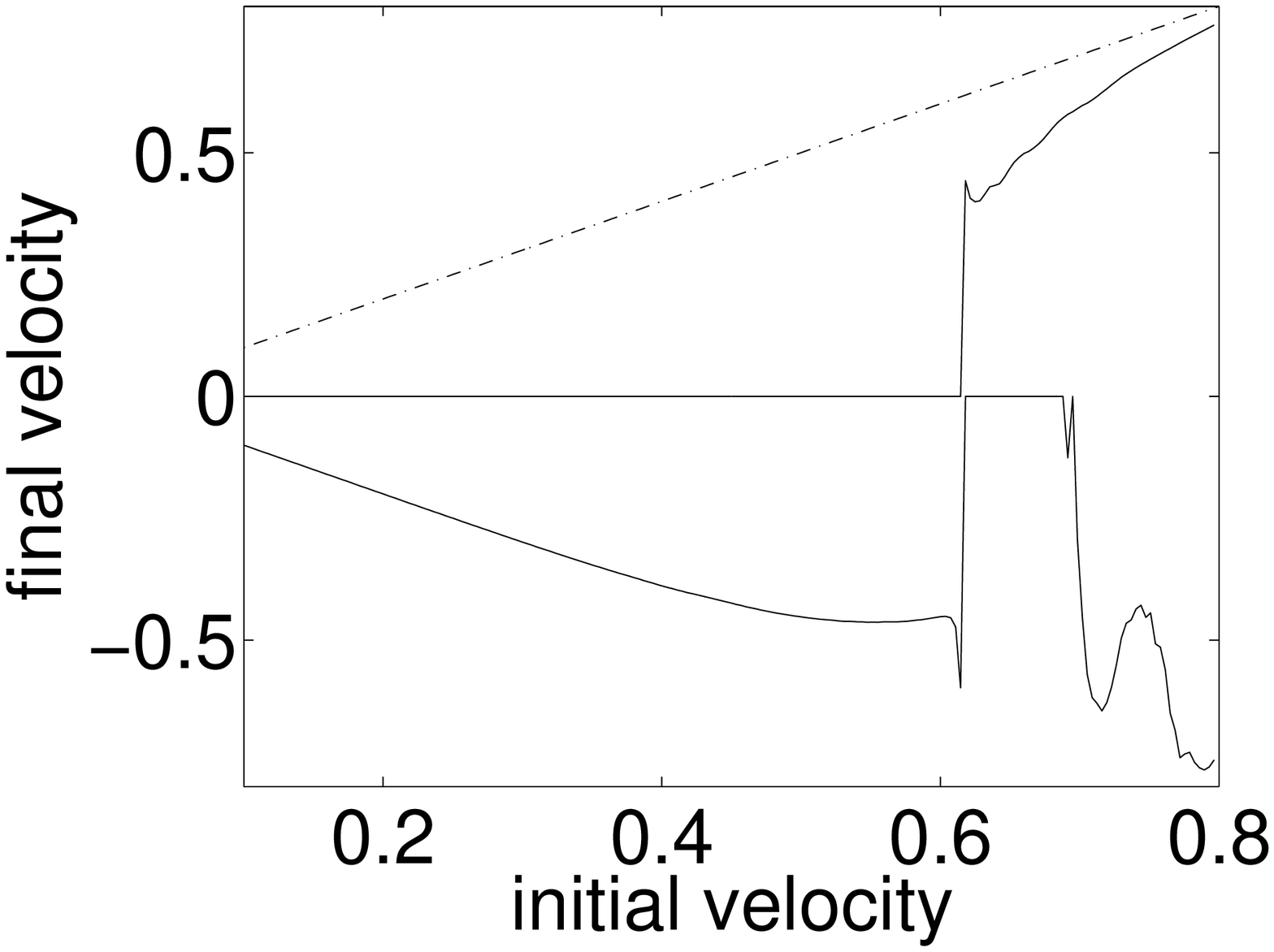}}
\caption{Initial velocity versus final velocity for a kink incident on
a narrow well without humps with $a=0$ and $b=\sqrt{10}/4.$
For the figure on the left, the well is empty whereas for the figure on 
the right, there is a kink in the well, which is initially at rest. 
\label{vivf2}}
\end{center}
\end{figure}

Figure \ref{vivf2} shows initial velocity versus final velocity for a 
narrow well without humps.  In the left figure, one kink travels towards 
the well with initial velocity $v.$  There is a ``resonant window'' 
at $v=0.22$ where back-reflection occurs. In comparison to Figure 
\ref{vivf1} the behaviour near the critical velocity is less abrupt.
For large initial speed, the initial velocity 
approaches the final velocity.  However, there is a region where the curve 
appears to ``wobble''.  In terms of our numerical simulations, we have 
checked sensitivity to stepsize in time and space independently. The 
wobble is more sensitive to the stepsize in space, but appears to be a 
genuine phenomenon. This could be a novel feature due to our potential.

The figure on the right in \ref{vivf2} shows initial velocity versus final 
velocity for the narrow well without humps, where a second kink is 
initially static in the well. At small to medium initial velocities, 
elastic and in-elastic back-reflection occurs. For velocity $v>0.6$ the 
static kink is ejected from the well and at marginally greater initial 
velocity, the first kink becomes trapped. For higher velocities both 
kinks escape the well in opposite directions. Again the final velocity of 
the first kink is not a monotonic function of the initial velocity.         

\section{Conclusion}
\label{Conclusion}

In this paper, we have discussed the interactions of kinks with wells and
barriers. We introduced a novel class of smooth potentials which included
pure wells and pure barriers of various heights and widths. These
potentials also give rise to wells with two side barriers and barriers
with two side wells, as well as double wells and volcano barriers.
The main advantage of the proposed potentials is that the static
kink-in-the-well solution is explicitly known. The asymptotic kink
solution for a kink located at very large $x_0$ is also known. Therefore,
a kink can be scattered off a kink-in-the-well and the only relevant
parameter is initial velocity of the incoming kink. When the exact
solution of the kink in the well is not known, then the trapped kink tends
to oscillate in the well, such that the scattering behaviour depends not
only on the velocity of the incoming kink but also on the phase of the
trapped kink, which is more difficult to control.

We studied the scattering of one kink off a well, mainly focussing on our
standard well with $\lambda(0)=\frac{1}{4}$ and two side humps, and
reproduced all the known phenomena such as trapping, back-reflection and
escape. We then compared numerically calculated critical velocities for
various barriers to an analytic approximation assuming elastic
scattering. As expected there was good agreement for pure barriers, and
less agreement for the volcano barrier and the barrier with side wells.
It would also be interesting to consider the two-kink system, with one 
kink initially static in a well and to compute the ``critical velocity'' 
of an incoming kink at which this static kink is knocked out of the well.    

The novel feature of this paper is the scattering of multi-kinks in the 
presence of a well. We calculated various trajectories for our standard 
well, such as elastic scattering, inelastic scattering, kink one replacing 
kink two in the well and a scattering where both kinks escape. 
Unfortunately, we were unable to find a double back-reflection for our 
standard well, but it is likely that such a trajectory exists for the 
right choice of well parameters and initial velocity. For the quartic 
well, we found double trapping of kinks and a novel type of 
back-reflection. We also plotted initial velocity
against final velocity, both for an empty well and a kink trapped in a
well.  In the case of our standard well with humps, we found that, for a 
single kink, the plot reproduced all the expected behaviour. 
In the case of an incoming kink and a kink initially in the well we found 
for small to medium initial velocities there is back-reflection that 
starts of as elastic then becomes inelastic. When the initial velocity 
becomes high, the static kink is ejected and there are two ``windows'' in 
which the initially moving kink is trapped. For a narrow well without 
humps the plot for a single kink revealed a ``resonant window'' and an 
interesting ``wobbling'' behaviour. The plots for two kinks showed that as 
the initial velocity increases, there is a point where the static kink is 
ejected and further on, the first kink becomes trapped. At high speeds 
both kinks escape the well, moving in opposite directions. Again, the 
final velocity of the first kink is not a monotonic function of the 
initial velocity.             

A long term aim of this line of work is to be able to design wells which
have desirable properties. In certain physical systems, it is possible to
make the coupling constant inhomogeneous, e.g. the ferromagnetic spin 
chain described in the introduction \cite{Wysin}. 
Barriers and wells might then be
used to control kink dynamics for example by acting as filters. An even
more ambitious aim is discussed in \cite{Corrigan:2004se} where a
particular kind of integrable point defect could be used to construct
simple logical gates.

Our approach is applicable to various different systems. The $\lambda
\phi^4$ kink allows similar wells and barriers which would enable the
study of kink anti-kink scattering in the presence of barriers and
wells. Soliton-well interaction has also been discussed for various other
solitons for example deformed sine-Gordon models \cite{AlAlawi:2008aa,
Bazeia:2007ua}, $\lambda \phi^4$ models \cite{AlAlawi:2007th}, Q-ball
systems \cite{AlAlawi:2009nr} or generalized sigma models 
\cite{Ferreira:2007ue} where our approach may again prove to be useful.

Reference \cite {Sutcliffe:1993wc} presents calculations for two 
interacting 
kinks using the method of Manton for constant $\lambda$.  In 
\cite{Piette:2006gw} a moduli space approximation was proposed which 
also takes account of the degrees of freedom of the well. See 
\cite{Kalbermann:1998pn} for related work on the $\lambda \phi^4$ kink.  
Our calculations imply that the an additional degree of freedom could be 
the slope of the soliton at the centre of the 
kink as parametrized by $a$ and $b$. One could therefore investigate 
moduli space dynamics applied to the model described in this paper, where 
$a,$ $b$ and $x_{0}$ for each kink become functions of $t,$ and the degree 
of freedom of the well is also taken into account. This is currently work 
in progress.

\section*{Acknowledgements}

SWG is grateful to Giota Adamopoulou for helpful conversations about
physical interpretations of this work. Thanks go to Andy Hone for
discussions and comments at various stages of the project. SK and RH
acknowledge the Nuffield Foundation for a Nuffield Science Bursary. SWG is
supported by the EPSRC and the SMSAS, University of Kent.

\begin{small}

\end{small}


\begin{thebibliography}{99}
\bibitem{Abdalla}
  E.~Abdalla, B.~Maroufia, B.~C.~Melgar and M.~B.~Sedra,
  ``Information transport by sine-Gordon solitons in microtubules,''
Phys.\ A\  Stat.Mech.Appl {\bf 301} (1984) 169-73.

\bibitem{Ablowitz:1991}
M.~J.~Ablowitz and P.~A.~Clarkson, {\it Solitons, Nonlinear Evolution
Equations and Inverse Scattering},  London Mathematical Society Lecture
Note Series vol.  149, Cambridge University Press, Cambridge (1991).

\bibitem{AlAlawi:2007th}
  J.~H.~Al-Alawi and W.~J.~Zakrzewski,
  ``Scattering of topological solitons on barriers and holes in two
$\lambda \phi^4$ Models,''
  J.\ Phys.\ A  {\bf 40} (2007) 11319
  [arXiv:0706.1014 [hep-th]].

\bibitem{AlAlawi:2008aa}
  J.~H.~Al-Alawi and W.~J.~Zakrzewski,
  ``Scattering of topological solitons on barriers and holes of deformed
  sine-Gordon models,''
  J.\ Phys.\ A  {\bf 41} (2008) 315206
  [arXiv:0802.1939 [hep-th]].

\bibitem{AlAlawi:2009nr}
  J.~H.~Al-Alawi and W.~J.~Zakrzewski,
  ``Q-ball scattering on barriers and holes in $1$ and $2$ spatial
Dimensions,''
  J.\ Phys.\ A  {\bf 42} (2009) 245201
  [arXiv:0902.4358 [math-ph]].

\bibitem{AlAlawi:2009rt}
  J.~H.~Al-Alawi,
  ``Collective coordinate approach to the dynamics of various
  soliton-obstruction systems,''
  [arXiv:0911.1804 [hep-th]].


\bibitem{Bazeia:2007ua}
  D.~Bazeia, L.~Losano, J.~M.~C.~Malbouisson and R.~Menezes,
  ``Classical behavior of deformed sine-Gordon models,''
  Physica D {\bf 237} (2008) 937
  [arXiv:0708.1740 [nlin.PS]].


\bibitem{Corrigan:2004se}
  E.~Corrigan and C.~Zambon,
  ``Aspects of sine-Gordon solitons, defects and gates,''
  J.\ Phys.\ A  {\bf 37} (2004) L471
  [arXiv:hep-th/0407199].

\bibitem{Fei}
  Z.~Fei, Y.~S.~Kivshar and L.~Vazquez,
  ``Resonant kink-impurity interactions in the sine-Gordon model,''
  Phys.\ Rev.\ A.\  {\bf 45} (1992) 6019-6030.

\bibitem{Ferreira:2007ue}
  L.~A.~Ferreira, B.~Piette and W.~J.~Zakrzewski,
  ``Dynamics of the topological structures in inhomogeneous media,''
  J.\ Phys.\ Conf.\ Ser.\  {\bf 128} (2008) 012027
  [arXiv:0709.3919 [hep-th]].

\bibitem{Ferreira:2007gu}
  L.~A.~Ferreira, B.~Piette and W.~J.~Zakrzewski,
  ``Wobbles and other kink-breather solutions of the Sine Gordon model,''
  Phys.\ Rev.\  E {\bf 77} (2008) 036613
  [arXiv:0708.1088 [hep-th]].

\bibitem{Goodman}
R.~H.~Goodman and R.~Haberman,
``Interaction of sine-Gordon kinks with defects: the two-bounce
resonance,''
Physica D 195, 303-323 (2004).

\bibitem{Javidan:2006js}
  K.~Javidan,
  ``Interaction of topological solitons with defects: Using a nontrivial
metric,''
  J.\ Phys.\ A  {\bf 39} (2006) 10565.

\bibitem{Josephson}
  B.~D.~Josephson,
  ``The discovery of tunnelling supercurrents,''
  Rev.\ Mod.\ Phys  {\bf 46} (1974) 251-4.

\bibitem{Kalbermann:1998pn}
  G.~Kalbermann,
  ``Trapped states and bound states of a soliton in a well,''
  [arXiv:hep-th/9805169].


\bibitem{Kalbermann:2004fc}
  G.~Kalbermann,
  ``The sine-Gordon wobble,''
J.\ Phys.\ A {\bf 37} (2004) 11603,
  [arXiv:cond-mat/0408198].


\bibitem{Manton:1981mp}
  N.~S.~Manton,
  ``A remark on the scattering of BPS monopoles,''
  Phys.\ Lett.\  B {\bf 110} (1982) 54.

\bibitem{Manton:2004tk}
  N.~S.~Manton and P.~Sutcliffe,
  {\it Topological Solitons,}
Cambridge, UK: University Press (2004).


\bibitem{McCall}
  S.~McCall and E.~L.~Hahn,
  ``Self-Induced Transparency,''
  Phys.\ Rev. {\bf 183} (1969) 457.

\bibitem{Perring:1962vs}
  J.~K.~Perring and T.~H.~R.~Skyrme,
  ``A model unified field equation,''
  Nucl.\ Phys.\  {\bf 31} (1962) 550.


\bibitem{Piette:2005wz}
  B.~Piette, W.~J.~Zakrzewski and J.~Brand,
  ``Scattering of topological solitons on holes and barriers,''
  J.\ Phys.\ A  {\bf 38} (2005) 10403
  [arXiv:hep-th/0508032].


\bibitem{Piette:2006gw}
  B.~Piette and W.~J.~Zakrzewski,
  ``Scattering of sine-Gordon kinks on potential wells,''
  J.\ Phys.\ A  {\bf 40} (2007) 5995
  [arXiv:hep-th/0611040].

\bibitem{Scott}
Alwyn Scott, {\it Nonlinear Science,}
Oxford: Oxford University Press (1999).

\bibitem{Sutcliffe:1993wc}
  P.~M.~Sutcliffe,
  ``Classical and quantum kink scattering,''
  Nucl.\ Phys.\  B {\bf 393} (1993) 211.

\bibitem{Wysin}
  G.~Wysin, A.~R.~Bishop and P.~Kumar,
  ``Soliton dynamics on an easy-plane ferromagnetic chain''
  J. Phys. C\  Solid State Phys. {\bf 17} (1984) 5975-91.

\end{thebibliography}
\end{document}